\documentclass[usegraphicx]{mn2e}

\newcommand\hi{{\sc Hi}}
\newcommand\etal{et\thinspace al.~}

\newcommand\zsol{\rm\,Z_\odot}
\newcommand\msol{\rm\,M_\odot}
\newcommand\ergs{{\rm\,erg\,s^{-1}}}
\newcommand\kms{{\rm\,km\,s^{-1}}}
\newcommand\zmin{z_{\rm min}}
\newcommand\zmax{z_{\rm max}}
\newcommand\fiii{F_{\rm III}}
\newcommand\barzero{\bar{z}_0}
\newcommand\barone{\bar{z}_1}
\newcommand\tcool{t_{\rm cool}}
\newcommand\rcool{r_{\rm cool}}
\newcommand\rturb{r_{\rm trb}}
\newcommand\vturb{v_{\rm trb}}
\newcommand\lturb{l_{\rm trb}}

\newcommand\rmix{r_{\rm mix}}

\def\spose#1{\hbox to 0pt{#1\hss}}
\def\simpropto{\mathrel{\spose{\raise 3pt\hbox{$\propto$}}
     \lower 3.0pt\hbox{$\sim$}}}

  \makeatletter
  \ifx\CUP@mtlplain@loaded\undefined  
  \makeatother  
    
  \else
    
  \fi

\loadboldmathitalic 
\loadboldgreek      
  
\makeatletter
\ifx\CUP@mtlplain@loaded\undefined  
  \newfont\bit{cmbxti10 at 9pt}
\else
  \newfont\bit{mtbxti10 at 9pt}
\fi
\makeatother  
 
\title[Population III stars]
{The number and metallicities of the most metal-poor stars}
  \author[M. S. Oey]  {M. S. Oey$^1$ \\
$^1$Lowell Observatory, 1400 W. Mars Hill Rd., Flagstaff, AZ, 86001, USA}
\date{Accepted 29 October 2002}
\pagerange{\pageref{firstpage}--\pageref{lastpage}}
\pubyear{    }

\def\LaTeX{L\kern-.36em\raise.3ex\hbox{a}\kern-.15em
    T\kern-.1667em\lower.7ex\hbox{E}\kern-.125emX}

\let\leqslant=\leq

\begin{document}

\label{firstpage}

\maketitle

\begin{abstract}

Simple, one-zone models for inhomogeneous chemical
evolution of the Galactic halo are used to predict the number fraction
of zero-metallicity, Population~III stars, which currently is empirically
estimated at $< 4\times 10^{-4}$.  These analytic models 
minimize the number of free parameters, highlighting the most fundamental
constraints on halo evolution.  There are disagreements of at least an
order of magnitude between observations and predictions in limiting
cases for both homogeneous Simple Model and Simple Inhomogeneous Model
(SIM).  Hence, this demonstrates a quantitative, unambiguous
discrepancy in the observed and expected fraction of Population~III stars.
We explore how the metallicity distribution of the parent enrichment
events $f(z_0)$ drives the SIM and predictions for the Population~III fraction.
The SIM shows that the previously-identified ``high halo'' and ``low
halo'' populations are consistent with a continuous evolutionary
progression, and therefore may not necessarily be physically distinct 
populations.  Possible evolutionary scenarios for halo evolution are
discussed within the SIM's simplistic one-zone paradigm.

The values of $z_0$ depend strongly on metal dispersal processes, thus
we investigate interstellar mixing and mass transport, for the first
time explicitly incorporating this into a semi-analytic chemical
evolution model.  Diffusion is found to be inefficient for 
all phases, including the hot phase, of the interstellar medium (ISM):
relevant diffusion lengths are 2 -- 4 orders of magnitude smaller than
corresponding length scales for turbulent mixing.  Rough relations for
dispersal processes are given for multiphase ISM.  These suggest that
the expected low-metallicity threshold above zero is consistent
with the currently observed limit.

\end{abstract}


\section{Introduction}

The search for zero-metallicity, Population~III stars provides one of the
most fundamental empirical constraints on the assembly of galaxies and
the process of chemical evolution.  In spite of decades of methodical
searches, to date no {\it bona fide} Population~III stars, {\it i.e.,}
having zero metallicity, have been identified (e.g., Beers 1999).
This lack is
often cited as problematic (e.g., Bond 1981; Cayrel 1996; Larson 1998):
a first generation of stars must have existed, and to date, the stellar
initial mass function (IMF) empirically has appeared rather robustly
universal.  We would therefore naively expect some surviving Population~III
stars today.  Yet, the on-going efforts of Beers {\etal}(e.g.,
1992; 1999) show a current upper limit to the fraction of Population~III 
stars in the Galactic halo of $< 4\times 10^{-4}$ (see \S 1.1 below).
However, there is surprisingly little quantitative discussion on the
expected relative fraction $\fiii$ of Population~III stars.  
Bond (1981) first emphasized the existence of a discrepancy in the
observed fraction of low-metallicity stars for the Galactic
halo population, in terms of the Simple Model for galactic chemical
evolution.  A few additional estimates for $\fiii$ have been made, and
these vary by orders of magnitude, ranging from $\fiii\lse
10^{-1}$ (Bond 1981) to $10^{-4}$ (Tsujimoto {\etal}1999).  A primary
objective of this paper is to further discuss the parameters that
determine $\fiii$ and make better-understood predictions.

Meanwhile, these searches for the most metal-poor stars in the Galaxy
have more clearly determined their metallicity distribution.
At present, the lowest-metallicity stars identified thus far have
logarithmic abundance relative to solar of [Fe/H] $\sim -4$.  There has
been a parallel effort obtaining abundance measurements of the
lowest-metallicity Lyman $\alpha$ absorption-line systems (e.g.,
Lu {\etal}1996; Songaila \& Cowie 1996).  
These are currently in the range of [Fe/H] $\sim -3$ (Lu {\etal}1996;
Prochaska \& Wolfe 2000; Ellison {\etal}2000).  
Pettini (2000) emphasizes that metals have been 
detected in {\it all} Ly $\alpha$ forest systems that have been
searched to date, including those with the lowest column densities
(Ellison et al. 2000; Songaila \& Cowie 1996).  It therefore has been
suggested that [Fe/H] in the range $-3$ to $-4$ represents a physical
minimum threshold in metallicity.  Indeed, it is widely thought that the
abundances of Galactic halo stars in this range exhibit the yields 
of only a few, even individual, supernovae (e.g., Audouze \& Silk 
1995; Ryan {\etal}1996; Shigeyama \& Tsujimoto 1998).

How many Population~III stars can we expect to find today?
Besides these, is there indeed a minimum metallicity threshold?  If
so, is its value of order [Fe/H]  $\sim -4$?  Predictions for these
parameters will necessarily have large uncertainties, since they
depend on poorly constrained relations between cosmic
galaxy evolution, star formation, and multiphase interstellar medium.
Nevertheless, some insight on these fundamental problems is possible
from simple arguments based on first principles and dominant effects.
These are re-examined here, with special emphasis on interstellar
element dispersal, and lead to new estimates of the expected numbers of
Population~III stars and the low-metallicity threshold.  The results
also offer insight on the star formation processes in the early Universe.

\subsection{Current empirical limit}

We first estimate the current limit on $\fiii$ for the Galactic halo.
Beers (1999) reports that his group's survey of metal-poor stars in
the solar neighborhood (e.g., Beers {\etal}1992; hereafter BPS) to date
finds at least 373 stars with [Fe/H] $<-2.5$.  However, none have been
found at zero metallicity, or even at [Fe/H] $\lse -4$.
Their survey is not designed as a complete survey of
the halo metallicity distribution function (MDF), thus we compare with
the halo MDF measured by Carney {\etal}(1996), who find 16\% of halo
stars to have [$m$/H] $<-2.5$.  Their measurements of [$m$/H], a
logarithmic metal abundance relative to solar, are calibrated to
correspond closely with [Fe/H] (Carney {\etal}1987).  
Thus, assuming that all the BPS stars with [Fe/H] $<-2.5$ are halo
members, we find that the metal-poor tail examined by the 
BPS survey implies a fraction of zero-metallicity stars $\fiii <
4\times 10^{-4}$.

\section{Predictions in the homogeneous limit}

What value of $\fiii$ should we expect?
The very simplest generalization is that, for a present-day mean metallicity
$\barone$, the number of generations of enrichment events is:
\begin{equation}\label{nmeans}
n = \barone / \barzero  \quad ,
\end{equation}
where $\barzero$ is the mean metallicity of these individual
enrichment events.  The present-day fraction of Population~III stars
is therefore $\fiii\sim 1/n$.  Thus, if we wish to interpret the
metal-poor empirical thresholds as estimates of $\barzero$, we can
set a lower limit on $\fiii$.  Taking
$\barzero\sim 10^{-5}\ \zsol$ therefore implies $n\sim 10^5$ to attain solar
metallicity.  Likewise, the Galactic halo, having a mean metallicity
around $0.1 \zsol$, requires $n\sim 10^4$, and hence
$\fiii\sim 10^{-4}$.  

However, this overly simplistic analysis does not account for the
consumption of gas to form stars, assuming a closed system.  We
therefore turn next to the Simple Model for homogeneous chemical
evolution (e.g., Schmidt 1963; Pagel \&
Patchett 1975; Tinsley 1980) to describe the relationship between $n$
and metallicity $z$.  The effect of gas consumption naturally implies
larger numbers of stars for earlier generations, thereby increasing
the predicted $\fiii$.  

As Bond (1981) demonstrated, the Simple
Model predicts a final metallicity distribution:
\begin{equation}\label{eqsimplefiii}
\phi(z) = y^{-1}\ \exp(-z/y) \quad .
\end{equation}
For yield $y\simeq z_1\simeq 0.02$, this implies that about 15\% of
Galactic halo stars should have [Fe/H] $\leq -2.5$.  At the time of
Bond's work, none of 129 halo objects, namely stars from the
Carney {\etal}(1979) proper-motion survey and halo globular clusters,
had been found to have [Fe/H] $< -2.6$.  Ten years later, Ryan \&
Norris (1991) reported that, within the errors, their newer dataset
did show agreement between the prediction and observations;
and another five years later, Carney {\etal}(1996) report 16\% of
stars in their expanded survey have [Fe/H] $< -2.5$, a value in
remarkable agreement with the prediction of equation~\ref{eqsimplefiii}!  

Is Bond's discrepancy resolved?  Let us now consider the number of
stars expected below the current observational limit of [Fe/H] $<
-4$.  Equation~\ref{eqsimplefiii} predicts a fraction $5\times
10^{-3}$ of halo stars to have metallicities below this value.
This is about an order of magnitude greater than the observed limit
found above:  about 12 stars should have been found by the BPS survey,
with assumptions in \S 1.1.  Thus it appears that Bond's
discrepancy has simply shifted to lower metallicity.  While our search
techniques are greatly improved, the saga of Bond's original
discrepancy is an important caveat against prematurely concluding that
the discrepancy exists. 

Therefore, the limit on observed zero-metallicity, i.e.,
Population~III, stars themselves offers perhaps a more powerful
constraint on the early chemical evolution and star formation
conditions.  However, it is difficult to use the Simple Model to
robustly predict the actual fraction of zero-metallicity stars:
equation~\ref{eqsimplefiii} shows that for low metallicities, the
predicted number of stars per unit $z$ is roughly constant.
Because complete homogeneity is assumed, the Simple Model cannot
accommodate the stochastic effects that dominate the first generations
of contamination.  We therefore require an inhomogeneous model for
chemical evolution that can account for the quantization and
stochastic nature of the earliest enrichment events. 

\section{Inhomogeneous limit: no mixing}

As mentioned above, the element abundances of the most metal-poor
stars are generally interpreted as reflecting the yields of the first
generations of stars.  The impetus is particularly provided by (McWilliam
{\etal}1995):  the marked change in abundance patterns for [Fe/H]$<
-2.4$, as is expected when these properties are determined
stochastically rather than statistically; and the large dispersions
in these properties, that are qualitatively consistent with stochastic
supernova (SN) yields. 
As emphasized by McWilliam {\etal}(1995) and Ryan {\etal}(1996), 
these abundance dispersions now {\it force} us to abandon the Simple
Model and consider inhomogeneous chemical evolution for the
lowest-metallicity populations.

In the limit of no ISM mixing, inhomogeneous chemical evolution can be
described in the simplest case as overlapping regions of contamination. 
Oey (2000; hereafter Paper~I) presents such a model.  This model is
essentially an inhomogeneous variation of the Simple Model, and offers
an opposing limit for the case of no homogenization.  We will refer to
this as the Simple Inhomogeneous Model (SIM), and briefly review some
of the basics here.  We note that Argast {\etal}(2000) present a
similar model that is numerically calculated; their study differs in
that they consider only individual SNRs, while Paper~I considers
contamination from multi-SN superbubbles.  In addition, the analytic
development in Paper~I more clearly shows the relevant
parameterizations that also apply to the numerical results.


\begin{figure*}
\includegraphics{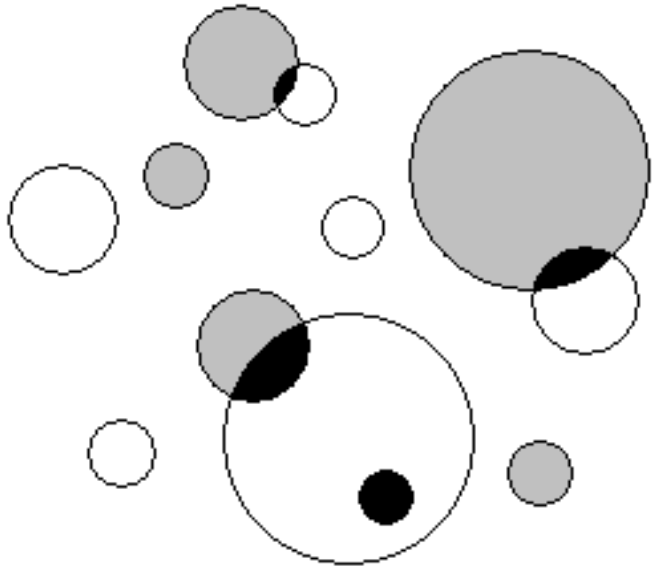}
\caption{ Schematic illustration of the Simple Inhomogeneous Model
described by Oey (2000), for $n=2$.  Each generation $k$ randomly
occupies filling factor $Q$.  Here, $k=1$ is shown by white regions
and $k=2$ by gray regions.  Overlapping regions for $j=2$ are shown in
black, while $j=1$ corresponds to both white and gray regions.
\label{overlap}}
\end{figure*}

For the metallicity distribution function of long-lived stars: 
\begin{equation}\label{eqinhomog}
N(z) = \frac{1}{n}\ \sum_{j=1}^{n}\ \sum_{k=j}^{n}\ D_{k-1}\ P_{j,k}\ N_j(z) 
	\quad ,
\end{equation}
where $N_j(z)$ is the MDF for stars born in areas resulting from $j$
overlapping polluted regions, $P_{j,k}$ is the probability of any
point having $j$ such overlapping regions after $k$ generations, and
$D_k = 1-k\delta$ is the fraction of gas remaining after $k$
generations.  This assumes that $\delta$, the fraction of original gas
consumed by each generation, is constant; thus it is related to the
present-day gas fraction $\mu_1$ by $\mu_1 = 1-n\delta$.  For any
given generation $k,\ P_{j}$ is given by the binomial distribution:
\begin{equation}
P_j = \Biggl({n \atop j}\Biggr)\ Q^j\ (1-Q)^{n-j} \ \ ,\quad
	1 \leq j \leq n \quad ,
\end{equation}
where $Q$ is the contamination filling factor for each generation,
which is assumed constant.  The corresponding probability of finding
zero-metallicity gas is: 
\begin{equation}
P_0 = (1 - Q)^n \quad .
\end{equation}
This model assumes instantaneous recycling, namely, that the products
of one generation immediately contaminate the following.
Figure~\ref{overlap} shows a schematic representation of this model.

As discussed more fully in Paper~I, the present-day mean metallicity for 
the SIM is characterized by $nQ$, the mean of the binomial distribution.
We assume that the metallicities for each generation of component
enrichment events are drawn from a fixed parent distribution $f(z_0)$;
these enrichment events are the units that build up the metal
abundance.  Here, they represent contamination by core-collapse 
SNe originating in OB associations and their superbubbles (see \S
3.2), although the model can accommodate $f(z_0)$ generated by
alternate mechanisms.

\begin{figure*}
\includegraphics{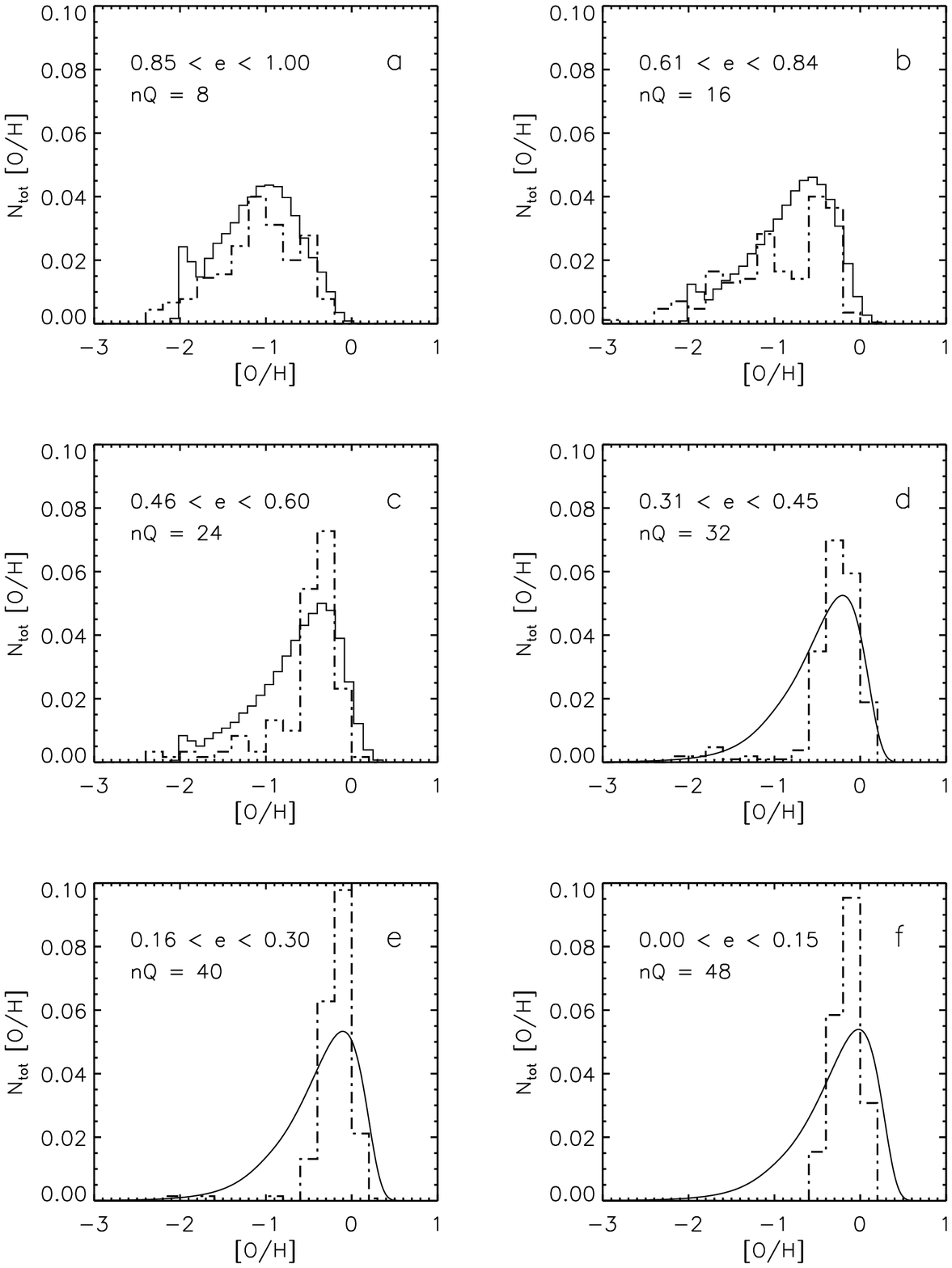}
\caption{Kinematic subsets of stars with different ranges in eccentricity $e$,
from Carney {\etal}(1996; dot-dashed lines), overplotted with Simple
Inhomogeneous Models having $nQ$ as shown (solid lines).  See text for
additional SIM parameters. 
\label{haloevol}}
\end{figure*}

For the Galactic halo, Paper I found that the SIM is in good agreement
with the observed MDF of long-lived stars for a model with $nQ\simeq
6.4$, range $\log\zmin$ to $\log\zmax$ of \break --3.7 to --2.0, and
present-day gas fraction $\mu_1 = 0$.  Here, we further examine 
kinematic subsets stars isolated by Carney {\etal}(1996; hereafter
CLLA96):  their Figures~5, 6, and 7 show MDFs for population subsets
defined according to kinematic criteria based on, respectively,
orbital velocity $V$, the quadratic radial and perpendicular velocity
$[U^2 + W^2]^{1/2}$, and eccentricity $e$.  These yield very
similar sequences of MDFs, so we arbitrarily select the sequence in
eccentricity (Figure~7 of CLLA96) for comparison with the SIM models.   
Figure~\ref{haloevol} shows the six subsets of data (dot-dashed
histograms), with the different ranges of $e$, increasing in steps of
roughly 0.15.  The MDFs are converted from [Fe/H] to [O/H] according
to the relation given by Pagel (1989):
\begin{equation}\label{eqPagel}
\rm [O/H] 
\left\{
\begin{array}{ll}
	 = \rm 0.5\ [Fe/H] ,
	& \rm [Fe/H] \geq -1.2 \\
\\
	= \rm [Fe/H] + 0.6 ,
	& \rm [Fe/H] < -1.2
\end{array}
\right.
\end{equation}
Overplotted in solid lines are SIM models with the same parameters
given above, but with $nQ$ increasing in steps of 8, as indicated.
These particular models take $Q = 0.8$, varying $n$ in multiples of 10.  

Figure~\ref{haloevol} demonstrates that this simple evolutionary
sequence shows important correspondences with the simple sequence in
eccentricity isolated by CLLA96:  (1) The qualitative shape
of the SIM MDFs, especially for the first three subsets in
Figures~\ref{haloevol}$a - c$, are in reasonable agreement with the
data.  (2) The relative rate in the shape change agrees well between
the models and data.  (3) The models and data agree reasonably well
quantitatively; in particular, the slowing rate of evolution in
the metallicities agree well between the models and data.
Note that there are important limitations for this SIM because of the
particular range for $f(z_0)$, which is discussed in \S\S 5 -- 6.
However, as shown in those sections, $f(z_0)$ and the corresponding
SIM can be adjusted.  

The point here is that these fundamental
correspondences strongly suggest that the stellar subsets represent an
evolutionary sequence, and that inhomogeneous evolution is necessary
to interpret the lowest-metallicity subsets.  CLLA96
suggest that these subsets represent a transition from ``high halo''
to ``low halo'' populations, such that the lower-eccentricity sets
represent a  disk or ``proto-disk'' population.  However,
the quantitatively simple progression of these eccentricity-defined
subsets as an evolutionary sequence could also suggest that these
stars are {\it all} members of a monolithic halo that formed slowly relative
to the contamination timescale.  Note that the simplistic bimodal formulation
of equation~\ref{eqPagel} may exaggerate the lack of low-metallicity
stars in Figures~\ref{haloevol}$d - f$; inspection of Figures~5 -- 7
of CLLA96 shows a more pronounced low-metallicity tail in [Fe/H].  But
if this apparent ``G-dwarf Problem'' in these subset MDFs is real, it
could be a manifestation of the disk G-dwarf Problem, as might be
expected if these populations are indeed related to the disk.

Regardless of the origin of these kinematic subsets, their identification
as an evolutionary sequence is strongly supportive of the SIM pattern
for chemical evolution.  The shape of the MDF at later evolutionary
stages seen in Figures~\ref{haloevol}$d - f$ is essentially
characteristic of homogeneous Simple Model.  However, at the earliest
stages, the SIM MDF shows a distinctly different shape that is
influenced by the parent $f(z_0)$ (see Paper~I and \S\S 5 -- 6 below).
This shape must then transition to that of the Simple Model.  {\it
That this is apparently seen in the kinematic subsets therefore
strongly suggests the SIM behavior and again supports the different
SIM interpretation of the halo MDF, in contrast to that of the Simple
Model} (see Paper~I and \S 6.3 below).  Thus we see that the SIM is
especially relevant to the lowest-metallicity regimes and hence it
is appropriate for investigating the following problems.  Note that we
therefore consider only core-collapse SN products with the instantaneous
recycling approximation.

\subsection{The fraction of Population III stars}

This inhomogeneous model provides straightforward predictions for the
fraction of 
Population~III stars, since the chemical evolution is characterized by
the contamination filling factor $Q$ and number of overlapping
generations $n$.  Note that in the case of the Simple Model, only the
very first generation of stars corresponds to Population~III; whereas
for the SIM, all regions with $j=1$ are included for
Population~III stars.  Hence, additional generations contribute,
provided that some of the star formation takes place in any remaining
pristine gas.  Thus, $\fiii$ is given by:
\begin{equation}\label{eqinhomogf3}
\fiii = \ \sum_{k=1}^n\ D_{k-1} P_{1,k} \biggl / \ 
	\sum_{j=1}^n \sum_{k=j}^n D_{k-1} P_{j,k} \quad ,
\end{equation}
where the numerator describes the probability of star formation
occurring in primordial gas over each generation $k$, and the denominator
corresponds to star formation at all degrees of contamination $j$ over all
generations (cf. equation~\ref{eqinhomog}).  In the $n=2$ example in
Figure~\ref{overlap}, Population~III corresponds to star formation in
the white and gray regions.  The predicted value of
$\fiii$ for a given MDF depends only on the product $nQ$ and is
highly robust ($\lse 5$\% variation) to different component $n$
and $Q$.  

For the halo model in Paper~I with $nQ = 6.4$,
numerical evaluation of equation~\ref{eqinhomogf3} 
gives a corresponding prediction of $\fiii = 0.28$ for these
parameters.  Paper I also shows excellent agreement for a SIM
of the Galactic bulge, using the same parameters as for the halo, but 
with a more evolved $nQ\simeq 44$; this yields $\fiii = 0.05$.  These
values for $\fiii$ are again startlingly large and emphasize a blatant
discrepancy with observations.

\subsection{The low-metallicity threshold}

In the SIM, the present-day mean metallicity $\barone$,
determined by $nQ$, depends on $f(z_0)$, which is characterized by its
mean, $\barzero$.  Paper~I estimated $f(z_0)\propto z^{-2}$,
within the limits\footnote{Note in Paper~I the values 
of $\log\zmin$ and $\log\zmax$ were inadvertently given for $m_y=1
\msol$ instead of the stated $m_y=10 \msol$; for the quoted value of
$m_y=10 \msol$, $\log\zmin$ and $\log\zmax$ are --3.7 and --2.0,
respectively.} $\log \zmin$ and $\log \zmax$ 
of --3.7 and --2.0, respectively, yielding a mean $\log \barzero=-3.1$.
These values were based on the simplistic assumption that metal yields of SNe
and their progenitors are uniformly dispersed within the associated
superbubble volumes, with a power-law distribution of $N_*^{-2}$ in
the number of SN progenitors $N_*$ per association (Oey \& Clarke 1998).
This is qualitatively consistent with observations by Hughes
{\etal}(1998) showing that smaller supernova remnants (SNRs) show
higher metallicities than larger ones.  The minimum $\zmin$
corresponds to metal dilution into the largest superbubbles, taken to
have radii $1300$ pc and gas density 0.5 cm$^{-3}$.  
However, $\zmin = -3.7$ corresponds to 
[Fe/H] $\simeq -2.6$, which is much larger than the observed minimum
threshold.  This is therefore a significant discrepancy between the
observations and the SIM having these parameters.

\section{Interstellar dispersal}

The above SIM model represents a limit in the case of no large-scale
mixing beyond the superbubble radii.  The assumption is
that metals are uniformly distributed within the
volumes of the hot superbubbles and simply cool {\it in situ},
remaining in place over timescale $\tau_n$.  The values of $z_0$ would be
reduced by dilution in the case of mixing, thereby possibly matching
the observed lowest metallicities.  Since the mean $\barzero$ is also
likely to decrease, a corresponding increase in $nQ$ would be required
to attain a given present-day metallicity.  This would therefore also
reduce the present fraction of Population~III stars.  What are
realistic lower limits for the values of $\zmin$ and $\barzero$?
The metallicities $f(z_0)$ of typical enrichment units depend on how far
the stellar products of one generation are spatially diluted within
the inter-generation timescale or duty cycle $\tau_n$.  

We now consider the interstellar mixing process in ordinary,
multiphase ISM.  Although this is 
difficult to constrain, some first attempts have been made by,
e.g., Bateman \& Larson (1993); Roy \& Kunth (1995); and Tenorio-Tagle
(1996).  Here, we examine the mixing process more closely.  We consider
that there are essentially two mass transport processes:
diffusion and turbulent mixing.  Roy \& Kunth (1995) name several
additional phenomena such as galactic rotation and shear, expanding
superbubbles, and hydrodynamic instabilities; however, these would all
contribute generally to turbulent mixing, and indeed are considered to
be the sources for interstellar turbulence.  Cloud kinematics and
collisions have also been discussed by Bateman \& Larson (1993) and
Roy \& Kunth (1995); however, these will only dominate if and when the
metal-enriched gas is entrained into cool clouds, and if the cloud
velocities exceed those of random ISM turbulent velocities.
Therefore, we will not consider cloud collisions here, and the
reader is referred to the mentioned studies for further discussion of that
process. 

\subsection{Diffusion}

Diffusion has been discussed in general terms by Tenorio-Tagle (1996),
who finds diffusion to be efficient for coronal gas typical of the hot
ionized medium (HIM), but inefficient for cooler gas such as the warm
ionized medium (WIM), warm neutral medium (WNM) or cold neutral medium
(CNM).  In order to quantitatively compare
the efficiency of diffusion with turbulent mixing, we evaluate
diffusion with the Maxwell-Chapman theory, in a more detailed examination
than the approximation used by Tenorio-Tagle.  

Following a point deposition, the concentration, or mass fraction of
the solute species, evolves under diffusion as, 
\begin{equation}
c_1(t) = \frac{M_1}{8\rho_2(\pi Dt)^{3/2}}\ 
	\exp\Bigl(\frac{-r^2}{4Dt}\Bigr) \quad ,
\end{equation}
where $r$ is distance from the origin point, $M_1$ is the total mass of
the solute, $\rho_2$ is the mass density of the ambient field particles,
and $D$ is the diffusion coefficient.  Following the formalism of
Woods (1993), the coefficient of mutual diffusion for two species of
different mass is,
\begin{equation}\label{eqD12}
D_{12} = \frac{3kT}{16Mn_0\Omega_D} \quad ,
\end{equation}
where $T$ is the thermal temperature; $M = m_1 m_2/(m_1+m_2)$ is the
particle reduced mass with $m_1$ and $m_2$ representing the masses of
the solute and field particles, respectively; $n_0 = n_1 + n_2$ is the
total number density, with $n_1$ and $n_2$ respresenting the solute
and field number densities, respectively; and $k$ is Boltzmann's
constant.  The particles are assumed to interact according to an
inverse power-law force $K_{12} d^{-\nu}$, where $d$ is the distance
between particles.

For ions, the particles repel under the Coulomb force with $\nu = 2$,
for which, 
\begin{equation}
K_{12} = Z_1 Z_2 e^2/(4\pi\epsilon_0) \quad ,
\end{equation}
where $Z_1$ and $Z_2$ are respectively the solute and field particle charges
in units of electronic charge $e$, and $\epsilon_0$ is the
permittivity of free space.  We also have,
\begin{equation}
\Omega_D = 2\pi\ {\rm ln}\ \Lambda\ \Biggl(\frac{kT}{2\pi M}\Biggr)^{1/2}
	\Biggl(\frac{K_{12}}{2kT}\Biggr)^2 \quad ,
\end{equation}
where $\Lambda$ is the ratio of the Debye length to the average
particle collision impact parameter: 
\begin{equation}
\Lambda = \Biggl(\frac{\epsilon_0 kT_e}{n_e e^2}\Biggr)^{1/2}
	\ \Biggl(\frac{\vert Z_1 Z_2\vert e^2}
		{4\pi\epsilon_0 M\overline{g^2}} \Biggr)^{-1} \quad .
\end{equation}
$T_e$ and $n_e$ are the electron temperature and density,
respectively; and $g$ is the relative speed between the solute and
field particles.  

For neutral particles, we can take $\nu= 5$, and (Jeans 1916):
\begin{equation}
K_{12} = kT(\nu - 1)\ \sigma^{\nu - 1} \quad ,
\end{equation}
where $\sigma$ is the distance between particle centers at collision
for the two species.  For atomic H and O interactions, $\sigma
\simeq 3.15\times 10^{-8}$ cm.  We also have (Woods 1993),
\begin{equation}
\Omega_D = \Biggl(\frac{\pi kT}{2M}\Biggr)^{1/2} 
	\Biggl(\frac{K_{12}}{2kT}\Biggr)^\frac{2}{\nu-1}\ A_1(\nu)\ 
	\Gamma\Bigl(3-\frac{2}{\nu-1}\Bigr) \quad ,
\end{equation}
where $A_1(5) = 0.422$ and $\Gamma(\frac{5}{2}) \simeq 1.329$. 

\begin{table*}
\begin{minipage}{100mm}
\footnotesize
\caption{Parameters for O diffusion in diffuse ISM \label{Tdiffusion}}
\begin{tabular}{@{}cccccc}
{Phase} & {$T$} & {$n$(H)} & 
	{O ion} & {log $D_{12}$} & 
	{$r_{\rm rms}\ ^a$} \\
{} & {(K)} & {(cm$^{-3}$)} & {} &
	{($\rm cm^2\ s^{-1}$)} & {(pc)} \smallskip\\
CNM & $1\times 10^2$ & 1.0 & O$^0$ & 18.25 & 0.06 \\
WNM & $8\times 10^3$ & 0.3 & O$^0$ & 20.72 & 1.0 \\
WIM & $1\times 10^4$ & 0.1 & O$^{+2}$ & 18.04 & 0.05 \\
HIM & $1\times 10^6$ & 0.003 & O$^{+5}$ & 23.64 & 30 \\
\end{tabular}

$^{\rm a}${Diffusion length for $1\times 10^8$ yr.}
\end{minipage}
\end{table*}

From equation~\ref{eqD12} we now evaluate the diffusion constants for
mutual diffusion of O and H in different interstellar phases, which are
given in Table~\ref{Tdiffusion}, along with the assumed temperature,
density, and O ionization stage.  The corresponding r.m.s.~diffusion
length for a timescale $t = 10^8$ yr, given by,
\begin{equation}
r_{\rm rms}^2 = \overline{r^2} = 6 D_{12} t \quad ,
\end{equation}
is also shown.  For small concentrations of the solute, $D_{12}$ is
insensitive to temperature gradients (Landau \& Lifshitz 1987), so we
do not consider a temperature differential between the two species.
Equation~\ref{eqD12} shows that $D_{12}$ is also insensitive to the
abundance of O, provided that $n_1\ll n_2$; here we take $n_1 =
10^{-3}\ n_2$.  We take $T_e$ and $n_e$ to be the same
as the thermal temperature and density for the H ions.

The values of $D_{12}$ shown in Table~\ref{Tdiffusion} are even
lower than those estimated by Tenorio-Tagle (1996), especially
when considering the generously low densities assumed here.  Most
significantly, Table~\ref{Tdiffusion} shows that even diffusion in the
HIM is much slower than previously suggested:  the scale length after
$10^8$ yr is only 30 pc, more than two orders of magnitude smaller than
estimated by Tenorio-Tagle.  Thus it appears that diffusion is
rarely efficient for mass transport.

\subsection{Turbulent mixing}

Therefore, it appears that turbulence must strongly dominate the
interstellar mixing process in all phases.  A mixing length scale is
given by Bateman \& Larson (1993):
\begin{equation}\label{eqBL}
\rturb = \bigl(2/3\ \vturb\ \lturb\ t\bigr)^{1/2} \quad ,
\end{equation}
where $\vturb$ and $\lturb$ are the characteristic turbulent velocity
and associated correlation length, respectively.  Determining an
appropriate value for $\lturb$ is problematic, since the \hi\ spatial
power spectrum in the ISM generally shows clean power-law slopes.
These have been studied in the Small Magellanic Cloud (Stanimirovi\'c
\& Lazarian 2001), Large Magellanic Cloud (Elmegreen {\etal}2001), and
the Milky Way (Dickey {\etal}2001) and found to be broadly consistent
with Kolmogorov turbulence over the scales considered here.  The
length scale most effective at  dispersing a coherent structure like a
superbubble would be that associated with the object.  Thus, if we
consider $\lturb= 50$ pc, over the same timescale $t=1\times 10^8$ yr,
we obtain: 
\begin{equation}\label{eqrturb}
\rturb = 58.3\ \bigl(\vturb/\kms\bigr)^{1/2}\ \rm pc \quad .
\end{equation}
As a rough first
estimate for $\vturb$, we take the soundspeed in the given ISM phase,
since this represents a characteristic velocity.  Thus
equation~\ref{eqrturb} shows that for $\vturb$ of order $1 - 100\
\kms,\ \rturb$ is 2 -- 4 orders of magnitude greater than the diffusion
lengths estimated in Table~\ref{Tdiffusion}.

\subsection{The Dispersal of metals}

We now consider the volume over which the metals produced within
superbubbles may be dispersed within the star-formation duty cycle
$\tau_n$.  Since the various ISM phases have different densities and
different associated soundspeeds, the effectiveness of turbulent
mixing depends strongly on the phase through which the material is
propagating.  This in turn depends on the cooling process.

Mac Low \& McCray (1988) considered the radiative cooling of
SN-driven superbubbles, deriving an expression for the cooling time
$t_{\rm rad}$ in terms of the input mechanical power $L_{38}$,
ambient number density $n_0$ and metallicity $\zeta$:
\begin{equation}\label{eqradcoolt}
t_{\rm rad} \simeq 16\ L_{38}^{3/11} n_0^{-8/11} \zeta^{-35/22}\ \rm Myr 
	\quad ,
\end{equation}
where $L_{38}$ and $\zeta$ are in units of $10^{38}\ \ergs$ and
$\zsol$, respectively.  They found that the radiative cooling time is
in general longer than the superbubble evolutionary timescale, for
solar metallicities.  Since we are considering
extremely low metallicity conditions, equation~\ref{eqradcoolt}
demonstrates that the first generations of superbubbles should all contain
hot, coronal gas after the final SNe explode:  for $\zeta=0.1,\ t_{\rm
rad}$ already increases by a factor of $\sim$40 from its duration at $\zeta=1$.

Heat transport processes are largely analogous to those for mass
transport, hence the above analysis implies that, for hot (10$^6$ K)
gas, the dominant cooling mechanism is turbulent mixing into the cool
ambient medium, rather than radiation.  As described by Landau \&
Lifshitz (1987), thermal homogenization by turbulence takes place 
through the mechanical mixing of fluid parcels.  Rapid energy
dissipation by thermal conduction ensures that 
temperature fluctuations on microscopic scales are always smaller than
those on larger scales.  Thus, the turbulent
mixing timescale for mass transport should also apply to that for heat
transfer.  However, if the HIM phase dominates the ISM
volume, then cooling by turbulent mixing may not be efficient (see
below). 

Clearly these turbulent processes are extremely difficult to
constrain.  We therefore make an attempt here to only estimate the
order of magnitude scale of these effects.  As a start, we consider
the cooling of material from HIM to WIM/WNM (hereafter WM).
We can roughly estimate the cooling time as the turbulent mixing time
for a quantity of hot gas mixing with 10 times that quantity of cool
gas, in number of particles.  This ratio implies cooling of the hot
gas to an equilibrium temperature that is of order $10^5$~K, from
which the plasma can efficiently cool further by collisional
ionization of He and H (Sutherland \& Dopita 1993), and metals if any  
are present.  We can estimate the turbulent mixing time from
equation~\ref{eqBL} for hot gas within a superbubble having a final
radius $R$, mixing with cooler ambient gas.  We take the density ratio
between the HIM and WM to be $n_1/n_2 = 0.01$, so for for mixing into ambient
WM, the relevant turbulent mixing time $\tcool$ is that required for
mixing the hot superbubble gas with  
one tenth its same volume of ambient WM, thus a total volume with an
equivalent radius $\rcool = 1.03 R$.  For hot gas mixing into 
CNM, the equivalent $\rcool = 1.003 R$.  More generally,
we can write, $\rcool = \epsilon R$.  As argued
above, the value of $\lturb$ most effective for mixing the
superbubbles should be similar to $R$.  This assumption is also
broadly consistent with the paradigm that superbubble kinematics are a
significant driver of ISM turbulence.  Setting $\lturb = R$
we thus obtain,
\begin{equation}
\tcool = \frac{3}{2}\frac{\epsilon^2 R}{\vturb} \quad .
\end{equation}
Again associating $\vturb$ with the soundspeed as above, this gives a
turbulent mixing/cooling time for hot gas:
\begin{eqnarray}
t_{\rm cool,W} & \simeq & 0.10\ \lturb(\rm pc)\ \ Myr\ \ \ (WM) \\
t_{\rm cool,C} & \simeq & 0.94\ \lturb(\rm pc)\ \ Myr\ \ \ (CNM) \ , 
\label{eqtcoolCNM}
\end{eqnarray}
for $\vturb = 16\ \kms$ and 1.6 $\kms$ for cooling into WM and CNM,
respectively. 

We now consider the dispersal of heavy elements, which initially are
contained by the hot gas within the parent superbubble.  If
the breakup of the shell allows immediate exposure of the
metal-enriched hot gas to the HIM, then turbulent mixing can occur in
this highly efficient phase until mixing with adjacent WIM and neutral
gas cools the hot material.  Since turbulence dominates both the
mass and heat transport, we can use limits set by turbulent mixing
parameters to make a rough first estimate of the dispersal length
scale for mixing within the HIM.  We first consider the fraction of
metal-enriched gas that has cooled after an interval $t$:
\begin{equation}
f_{\rm cool} = \frac{\rturb^3}{(\epsilon_2 R)^3} = 
	\Bigl(\frac{2}{3} v_2\Bigr)^{3/2}
	\epsilon_2^{-3} R^{-3/2} t^{3/2} \quad ,
\end{equation}
where $v_2$ and $\epsilon_2$ are parameters for WM or CNM,
depending on the assumed ambient medium.
The remainder is still hot and can reach a distance $r_{\rm trb}$
according to equation~\ref{eqBL}, with $\lturb = R$ as before.  The total
volume ultimately occupied by dispersing hot gas is therefore, 
\begin{equation}
V_1 =  \int_0^{\rcool}\ (1-f_{\rm cool})\ 4\pi\rturb^2\ d\rturb \quad .
\end{equation}
Integrating this, we find the equivalent length scale $(3V_1/4\pi)^{1/3}$:
\begin{equation}\label{rmixturb}
r_{\rm mix}
	= 2^{-1/3} \epsilon_2
	\Bigl(\frac{v_1}{v_2}\Bigr)^{1/2}\ R \quad ,
\end{equation}
where subscripts 1 and 2 denote parameters for the dispersing and
ambient media, respectively.  For a density ratio $n_1/n_2 = 0.01$ and
$v_1/v_2 = 10$, as we assume for HIM ($v_2 = 160\ \kms$) mixing into
WM, equation~\ref{rmixturb} reduces to $r_{\rm mix} = 2.6\ R$ for the
characteristic mixing distance.

The extent of mixing within the WM itself is difficult to estimate
analytically.  For a continuous WM, once the hot gas has cooled
to this phase, individual particles could in principle be transported
within this phase during long time scales.  Conversely, cooling in
this phase takes place rapidly if its global, or local, heating sources are 
not maintained.  However, it again seems likely that turbulence
determines the circumstances that permit individual gas parcels to
cool from this phase.
Considering now the analogous dispersal of WM into CNM, 
we again have $n_1/n_2\simeq 0.01$ and $\epsilon_2 = 1.03$, therefore
also resulting in equation~\ref{rmixturb}.  We adopt the original value of $R$;
while the geometric distribution for the cooled superbubble gas is
greater than $R$, the cooled material is now at the WM density and its
equivalent volume is reduced by a factor of 0.01 for the adopted densities.
Since $R$ is the characteristic length, it remains a reasonable
length scale for equation~\ref{rmixturb} within the large
uncertainties of this approximation, and also helps offset the likely
additional time or distance distance needed to encounter CNM.

Thus, in a first crude estimate, we take the efficiency of element
dispersal in the WM to be the same as in the HIM.  Although turbulent
mixing itself is less efficient, this is counteracted by the likely longer
residence time within this phase if a continuous WM exists, as
it does in local star-forming galaxies.

Meanwhile, mixing also takes place in the CNM.  Assuming
mixing of hot superbubble gas with ambient CNM starts at 
$t_f$, the time at which the shell attains its final, pressure-confined
radius, equation~\ref{eqBL} gives a length scale for mass transport
within the CNM:
\begin{equation}\label{rmixCNM}
\rmix = \Bigl[\frac{2}{3} v_2 R\ 
	(\tau_n - t_{\rm cool,C} - t_f)\Bigr]^{1/2} \quad ,
\end{equation}
where $v_2$ is now the turbulent velocity for the CNM.  Oey \& Clarke
(1997) give for $t_f$:
\begin{equation}
t_f = t_e\ R/R_e \quad ,
\end{equation}
where $t_e \simeq 40$ Myr is the duration of SN mechanical power for the
superbubble, and  
\begin{equation}
R_e = \frac{5}{7^{1/2}}\ \rho_2^{-1/2}\ P_2^{1/2}\ t_e \quad ,
\end{equation}
with $P_2$ as the ambient total pressure.  

It is again essential to bear in mind that these estimates are crude
approximations, with transport within the WM especially
uncertain.  Full hydrodynamical simulations are necessary to more
reliably constrain the mixing processes.  However, it is encouraging
that first attempts at numerical results are broadly consistent with the
above estimates (de Avillez \& Mac Low 2002).  In general, the parameters
adopted above should tend toward upper limits on the dispersal length;
for example, adopting the soundspeed for $\vturb$ is generous, and the
influence of magnetic fields will generally inhibit mixing.

\begin{figure*}
\includegraphics{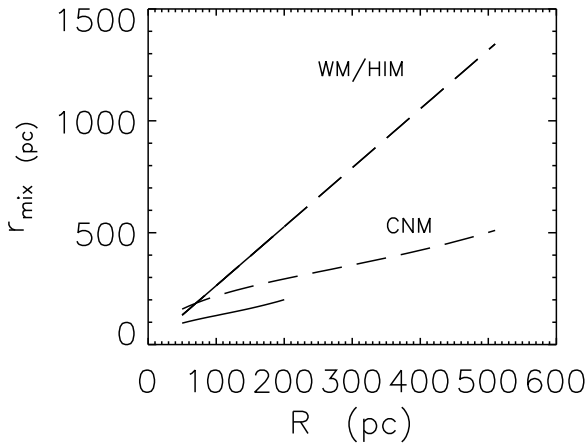}
\caption{Estimates for the dispersal length $\rmix$ from
superbubbles having final radii $R$.
The limiting cases correspond to dispersal of hot superbubble gas into
HIM and/or WM, and CNM as indicated; dashed lines show
models with $\tau_n = 500$ Myr and solid lines show $\tau = 200$ Myr.  
\label{fphasemix}}
\end{figure*}

Figure~\ref{fphasemix} shows the dispersal length $\rmix$ as a
function of superbubble final radius $R$ as given by
equations~\ref{rmixturb} and \ref{rmixCNM}.  The HIM/WM dispersal
shows a linear relation between $\rmix$ and 
$R$ with slope 2.6 (upper set of lines), as determined above, and the
slope for the CNM limit  is $\sim$0.7 (lower set of lines).  
These represent crude limits to the dispersal of elements by
turbulence.  Estimates are shown for star formation duty cycle times
of $\tau_n = 500$ Myr (dashed lines) and 200 Myr (solid lines).
These values of $\tau_n$ limit the largest $R$ to roughly 500 pc and 200
pc, respectively, as indicated by equation~\ref{eqtcoolCNM}; larger
objects may exist, but will not be able to cool within time $\tau_n$.
Thus, for $\tau_n$ characteristic over the relevant period of star
formation, the superbubbles have a corresponding range in $R$.

It is apparent that the dispersal length is sensitive to the ISM phase 
balance, with the slope of the relation between $\rmix$ and $R$
ranging from about 0.7 to 2.6.  For normal or low star formation
rates, this suggests that the dispersal scale is a
factor of up to 2.6 times the parent superbubble radius, or a factor
of up to $\sim$20 times the parent superbubble volume:  i.e., roughly an
order of magnitude for this crude analysis.

Beware that we assume that cooling of the hot superbubble gas by
turbulent mixing determines the HIM residence time.  Estimates for
most nearby galaxies having normal star-formation rates show low HIM
filling factors (e.g., Oey {\etal}2001).  However, if the HIM
filling factor is large, then the newly-produced metals could be
transported indefinitely.  Mixing would
take place extremely efficiently in a HIM-dominated ISM, justifying
instantaneous mixing approximations in chemical evolution models.  On
the other hand, a HIM-dominated ISM probably implies galactic
superwinds, complicating the fate of the stellar products.
Indeed, a blow-out or galactic superwind
would return us to the conventional problem of losing an unknown
amount of metals from the system.

\section{Inhomogeneous limit:  dispersal by mixing}

In short, the preceding section gives crude arguments that suggest
that the dilution of metals beyond the parent superbubble radii is
roughly another order of magnitude by volume.  Note that if the hot,
metal-bearing gas is vented out of a gaseous galactic disk, that these
arguments present an upper limit to extended dilution within the
disk where subsequent stars presumably form.  Moreover, merging
between adjacent contaminated regions is greatly enhanced by the
mixing process, and limits the dilution of metals within a galaxy's
star forming volume. 

We now return to the Simple Inhomogeneous Models, bearing in mind that
these represent a limit for the case of no homogenization, i.e.,
mixing is limited only to within the $\rmix$ argued above.  We can
reexamine the values of $\zmin$ and $\barzero$ of individual
enrichment events, as well as enrichment by the first generation
of stars.  In \S 3.2, we had $\log\zmin = -3.7$ and $\log \barzero
\sim -3.1$, based on the range of $z_0$ estimated in Paper~I for no
interstellar mixing beyond the superbubble radii.  Since the
relationship between $\rmix$ and $R$ is virtually linear, we can
simply scale the expected values of $z_0$ from the no-mixing values,
thus decreasing them by up to an order of magnitude, as estimated
above.  Note that if the superbubble filling factor $Q$ is high
enough, mixing from adjacent regions will overlap, so this is again a
conservative estimate in maximizing the decrease in $\barzero$.

\begin{figure*}
\includegraphics{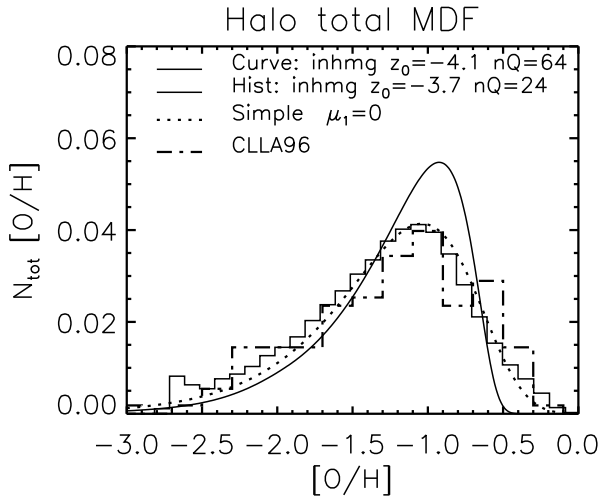}
\caption{Observed Galactic halo MDF from Carney {\etal}(1996;
dot-dashed line) overplotted with the inhomogeneous models for mixed
dispersal.  The solid histogram shows the low-evolution model
($\barzero = -3.7,\ nQ = 24$), and the solid curve shows the high-evolution,
model ($\barzero = -4.1,\ nQ = 64$).  The Simple Model is also shown,
with the dotted line. 
\label{fhalomods}}
\end{figure*}

\subsection{Mixed-dispersal, high-evolution SIM}

Scaling the original range in $z_0$ by a factor of 10 gives
$-4.7 < \log z_0 < -3.0$ and $\log \barzero \sim -4.1$.
For most chemical evolution models, the reduction in $\barzero$
implies that the number of enrichment generations $n$ must be
increased by roughly the same factor to achieve a given present mean
metallicity.  With the new range in $z_0$, the SIM
in \S 3 fits the Galactic halo MDF with $nQ = 64$, now predicting
$\fiii\sim 3\times 10^{-2}$.  This is an order of magnitude smaller
than $\fiii \sim 3\times 10^{-1}$ predicted by the no-mixing SIM
in \S 3.1.  However, an
important consequence of reducing $\barzero$ here is that the 
shape of the predicted MDF no longer agrees as well with the
observations as the original model with $\log \barzero = -3.1$.
The dot-dashed histogram in Figure~\ref{fhalomods} shows the observed
halo MDF from Carney {\etal}(1996), using their field star sample of
135 stars having retrograde velocities ($V < -220\ \kms$).  The data
are converted to [O/H] from [Fe/H] according to equation~\ref{eqPagel}.
The evolved SIM with $nQ = 64$ is
overplotted with the solid curve.  We have modified the conversion 
from $z$ to [O/H] used in Paper~I:  where previously we simply used
[O/H] $ =\log z + 1.7$, we now adopt a more exact relation, [O/H] $=
\log[Z/(24.32 - 128\thinspace Z)] + 3.07$.  The metallicity unit in
the SIM is the mass ratio of metals to H and He, thus $z = Z/(1-Z)$,
where $Z$ is the conventional mass fraction of metals.  The SIMs
assume a default SN yield of 10 $\msol$ and present-day gas fraction
$\mu_1 = 0$. 

Note again that this high-evolution case accounts for the complete dilution
of metals from each OB association, but it does not account for mixing {\it
between} the discrete enrichment regions, which corresponds to
homogenization.  As discussed in Paper~I, such 
homogenization would cause the models simply to tend toward the Simple
Model; hence in principle, the true distributions should be bounded by
the Simple Model and the SIMs.  Figure~\ref{fhalomods} also shows the
Simple Model for $\mu_1 = 0$ (dotted line).  

\subsection{Mixed-dispersal, low-evolution model}

In \S 3.2 we saw that the low-metallicity tail for the 
no-mixing model does not extend to metallicities as low as those observed.
(Note that although in Figure~\ref{fhalomods} the CLLA96
data are not shown to extend below [O/H] $\sim -3$, the current
observed limit does extend to the equivalent of [O/H] $\sim -3.4$.)
We can force the SIM to match the observations by
modifying the distribution $f(z_0)$, extending only $\zmin$
by a factor of 10, so that $-4.7 < \log z_0 < -2.0$.  The upper limit
represents contamination from single SNe, thus individual yields and
stochastic mixing effects could plausibly maintain a high $\zmax$.
Keeping the other input parameters the same, it remains possible to
match the observed halo MDF.  Owing to lack of computational
resolution, we cannot run our exact numerical models beyond $nQ \simeq
30$, but the behavior of the allowed models shows that $nQ$ now needs
to be evolved up to $nQ\sim50$ to match the data.  In lieu of this
model using the full required range of $z_0$, Figure~\ref{fhalomods}
shows a similar model with a range $-4.4 < \log z_0 < -2.0$ and $nQ =
24$ (solid histogram).  Comparison of this model with the no-mixing model in 
Figure~2 of Paper~I clearly shows that extending the lower range of
$z_0$ will allow an excellent fit to the observed halo MDF, including
the low-metallicity threshold.  This modification may also be applied
to the modeled halo subsets shown in Figure~\ref{haloevol} (again,
computational limitations prevent our explicit modeling of these
subsets). 

The high-metallicity tail of the MDF for this model, as for the
no-mixing model, reflects the remnant power-law distribution of
$f(z_0)$ (Paper~I); whereas for the model presented in the previous
section, the MDF has evolved to sufficiently high 
metallicities that the form of $f(z_0)$ is no longer manifest.  We
therefore refer to the former as ``low-evolution'' models
and to the latter as a ``high-evolution'' model.  Although, for the
case with extended range in $z_0$, the evolutionary parameter $nQ\sim
50$ is numerically similar to that in the high-evolution model, we
still designate this as a low-evolution model, since the tail of the
parent $f(z_0)$ is still apparent.  For the
low-evolution models, equation~\ref{eqinhomog} must be evaluated
numerically, whereas the high-evolution models can be computed with an
analytic approximation (Paper~I).  

The agreement between the data and the SIM does depend
on $f(z_0)$ ranging up to roughly $\log z_0 \sim -2.0$, which
corresponds to [Fe/H] $\sim -0.6$.  This causes the high-metallicity 
tail of $f(z_0)$ to remain manifest until the mean $z$ evolves to 
comparable values (see Paper~I).  Thus the assumed range in $z_0$
spans almost three orders of magnitude.  This range is large, but it
is unclear whether it is implausible (see \S 6.4 below); the strength
of this model is primarily that it matches the data.  With $nQ\sim
50$, this model predicts $\fiii\sim 4\times 10^{-2}$.  

\section{Discussion}

\subsection{The low-metallicity threshold}

The observed low-metallicity threshold, if it is real, provides an
important constraint on the models by limiting $\zmin$.  As we saw
above, the model MDFs, evolutionary state,
and predicted $\fiii$ are all sensitive to the form and range of
$f(z_0)$, especially $\zmin$.  Fields {\etal}(2002) analytically
demonstrate the interrelation between $\zmin$, the metallicity
dispersion, and the present-day metallicity.
Besides dilution by mixing, another model parameter that affects
$\zmin$ is the SN yield $m_y$.  The SIMs
assumed $m_y = 10\ \msol$, which could be overestimated by up to an
order of magnitude.  Reducing $m_y$ to $1\ \msol$ would
simply reduce the values of $z_0$, hence $\barzero$, by another order of
magnitude.  For the mixed-dispersal models, the estimated range of
the threshold $\log \zmin$ thus becomes roughly --5.7 to --4.7,
corresponding to [Fe/H] $\sim -4.6$ to --3.6.  This agrees well with
the observed lower limit $-4 <$ [Fe/H]. 

The lowest expected stellar metallicities could conceivably be below
the characteristic $\zmin$ if $\zmin$ varies over time.  Although these
SIMs do not accommodate an evolution of
$f(z_0)$, we briefly consider mixing processes for the earliest
conditions of star formation.  The very first generation of stars
should take place in WNM since the lack of metals in 
the early Universe maintains temperatures around 500 -- 1000 K at ISM
densities (e.g., Barkana \& Loeb 2001).  As subsequent generations of
stars begin to generate WIM, the production of metals soon enables
cooling and the formation of a CNM phase.  But for the first
generation, metal-enriched material from the first SNe could mix for
an extended period of time, since cooling below 
WNM temperatures is difficult.  This is also likely to lengthen the
timescale for forming the next generation of stars, to a value larger
than the characteristic $\tau_n$ at later times.  These factors argue
for widespread dispersal 
of metals produced by the first generation.  On the other hand, the
character of interstellar turbulence should be different from that in
the conventional multiphase ISM:  since there is no pre-existing
stellar mechanical feedback, turbulent mixing may be substantially less
efficient in dispersing the first stellar products.  As seen in \S
4.1, the competing process of diffusion is a much less effective
mixing agent than turbulence.  It is therefore difficult to constrain
the ISM metallicities resulting from the first generation of stars.
They could be either significantly lower than $\barzero$, or, in the
case of poor mixing, higher than $\barzero$.  At this stage, the value
of $\zmin$, taken as a characteristic value over all stellar
generations, is therefore about the best estimate we can make for the low
metallicity threshold above primordial.  As shown above, the rough
estimates of $\zmin$ are compatible with the currently observed limit.

We also note that other mechanisms for a low-metallicity threshold
have been proposed, that do not rely on the existence of an intrinsic
threshold in the parent enrichment units.  For example, Hernandez \&
Ferrara (2001) estimate an effective low-metallicity threshold around
$\log\zmin/\zsol \sim -4.3$; they obtain this value by estimating the number
of progenitor sub-halos needed to assemble the Galaxy, and extrapolate
$\zmin$ as the mean cosmic abundance corresponding to individual such
sub-halo at the time of its formation.  Schneider {\etal}(2002)
propose that a similar $\log\zmin/\zsol\sim -4$ is a physical threshold
needed to form low-mass ($\sim 1 \msol$) stars, with pre-enrichment
provided by a first generation of only supermassive stars.  
The value $\log\zmin/\zsol$ of --4.0 corresponds to [Fe/H] of --4.6, thus
these estimates are quite similar to the value estimated here,
with large uncertainties admitted by all these studies.  Note that 
these alternative approaches avoid addressing the parent $f(Z)$, which is the
straightforward origin of $\zmin$ estimated in this work.

In the above analyses, including our own and those just described,
$\zmin$ is not necessarily a hard limit, but a characteristic value
for the lowest expected metallicities; stochastically, it is likely
that some stars will be found at metallicities anywhere between zero
and $\zmin$, depending on their specific parent interstellar mixing
conditions.  However, $\zmin$ should correspond to a sharp drop-off in
observed MDFs. 

\subsection{The fraction of Population~III stars}

Besides the Simple Model analysis by Bond (1981), there are a couple
other predictions in the literature for $\fiii$, the fraction of
zero-metallicity, Population~III stars.  Cayrel (1996) makes a rough
estimate of $\fiii\sim 10^{-5}$ for the entire Galaxy.  This prediction
is based on SN contamination within individual clouds, and the author
emphasizes the uncertainty caused by the arbitrarily assumed number of
these clouds.  Tsujimoto {\etal}(1999) also make a prediction based on
a detailed model of single SN contamination within clouds, with more
complex chemical 
regulation determined by the triggering of subsequent star formation
in SNR shells.  They predict $\fiii\sim 10^{-3} - 10^{-4}$ for the
Galactic halo.  However, it is unlikely that the occurrence of star
formation is dominated exclusively by triggering in SNR shells, and the 
action of multi-SN superbubbles causes substantial differences in ISM
dynamics from the assumed individual SNe (e.g., Oey \& Clarke 1997). 
Here, we emphasize different parameters, in particular $f(Z)$, thereby
presenting a complementary perspective on the dominant processes.

As can be seen above, our Galactic halo predictions for $\fiii$
are presented in simpler terms, and eliminate most of the difficult
parameters used by these other authors.  Thus, the contrast with the
empirical limit on $\fiii$ should yield better-understood constraints
on early star formation and enrichment.  To summarize:  the
homogeneous, Simple Model predicts a fraction $5\times 10^{-3}$ for
stars having [Fe/H] $< -4$; with the no-mixing SIM, we predicted 
$\fiii\sim 2\times 10^{-1}$; considering metal dispersal by mixing, we
obtain a high-evolution SIM and a low-evolution SIM, both
predicting $\fiii\sim 3 - 4\times 10^{-2}$.  These predictions are all
1 -- 3 orders of magnitude greater than the observed upper limits.

The SIM predictions are based on parent
enrichment events whose metallicity distribution $f(z_0)\propto z^{-2}$ is
derived from dilution of SN products into volumes that are referenced
to their SN-driven superbubbles (\S 3.2 above; Paper~I).  Although,
as emphasized above, the mixing process is rather uncertain, it is
apparent that regions of contamination will ultimately begin mixing
together and homogenizing, especially if mixing is even more efficient
than the generous estimate 
determined in \S 4.3.  Thus this condition imposes a likely limit to
the minimum $\barzero$ driven by dilution of parent enrichment
events.  Once the contaminated regions begin homogenizing, the
system should approach the homogeneous limit of the Simple Model.

We have seen above that $\fiii$ depends rather strongly on $f(z_0)$.
Although we used metal dilution within superbubbles to derive
$f(z_0)\propto z^{-2}$, it seems fairly easy to obtain a similar power-law
distribution in $z_0$ via other processes.  Thus these particular
SIMs are fairly robust and 
{\it independent of the actual origin of the $f(z_0)$ distribution.}
Note also that, although we have differentiated the SIMs
as ``no-mixing'' and ``mixed-dispersal,'' the only technical
difference between them is in the range of $z_0$.
As seen in \S 5, reducing the value of $\barzero$ 
implies a reduction in $\fiii$ by roughly the same amount.  Thus if we
invoke an overestimate in the default SN yield $m_y$ by a factor of
10, and take $f(z_0)$ for the high-evolution, mixed-dispersal model,
we obtain minimum estimates for $z_0$ of $-5.7 < \log z_0 < -4.0$.
This decreases the predicted $\fiii$ for that model by another
factor of 10, to $\fiii\sim 10^{-3}$ , which is still an order of
magnitude larger than the current empirical limit.

Besides $f(z_0)$, the SIMs for a given $nQ$ depend
only on the present-day gas fraction $\mu_1$, which was assumed
to be $\mu_1 = 0$.  Increasing $\mu_1$ will decrease the number of
contaminating generations required to attain the present-day
metallicity, hence would only serve to increase $\fiii$.

Therefore, given that the predictions and observations disagree for
both limiting cases, the homogeneous Simple Model and the SIMs,
it seems clear that {\it we can finally demonstrate a quantitative,
unambiguous discrepancy in the observed and expected 
numbers of Population~III stars,} based on our stochastic,
inhomogeneous expansion of the Simple Model.

\subsection{Predicted and observed MDFs}

We now consider the predicted and observed MDF's in their entirety.
The Simple Model and the no-mixing SIM were discussed 
in Paper~I.  The high-metallicity turnover for the homogeneous,
Simple Model results purely from the consumption of gas to form stars.
In contrast, although the MDFs are similar, the high-metallicity tail in
the low-evolution SIMs reflects the tail of $f(z_0)$.
On the low-metallicity side, Figure~\ref{fhalomods} shows that
the SIMs have distributions similar to the Simple Model.  This is
expected, since the SIMs approach the Simple Model
distribution in this regime (Paper~I).  However, unlike the Simple
Model, the SIMs have a minimum threshold metallicity
set by $\zmin$.  As determined in \S 6.1, the predicted
low-metallicity threshold is not inconsistent with that suggested by
the data.  

Since the Simple Model has no such lower limit, it
therefore predicts too many stars below the observed threshold.
But aside from this discrepancy, there is good agreement between the halo
data and Simple Model.  This could therefore 
suggest complete homogenization, although paradoxically, the similar
agreement for the low-evolution SIMs likewise suggests poor
homogenization.  As discussed above, the mixed-dispersal,
low-evolution SIM can 
produce a minimum $z$ that matches the observed threshold, as well
as match the observed halo MDF.  It therefore best reproduces the
data, out of the four models.  

Figure~\ref{fhalomods} shows that the high-evolution
SIM does not agree with the observed halo MDF, particularly for the
high-metallicity tail.  As demonstrated in Paper~I, 
the high-metallicity tail in this model continues to drop off more
steeply as the system evolves.  The disagreement implies that this
model does not properly account for the dominant effects.  The
steep, high-metallicity drop-off characterizes an
evolved system, as dictated by $nQ$ and $\barzero$.  Hence,
a low-evolution model may be a better representation of the system.
Alternatively, the halo may be better described by fully homogenized,
rather than inhomogeneous chemical evolution, thereby implying that
the Simple Model best represents the system, with the caveat that the
lowest metallicities are not observed. 

\subsection{Halo evolution}

Both the homogeneous and inhomogenous models considered here are
essentially one-zone models, therefore implying a single coherent
formation process for the Galactic halo.  This is more consistent with
the monolithic collapse model of Eggen, Lynden-Bell, \& Sandage
(1962), than the more favored Searle \& Zinn (1978) multi-fragment
assembly model.  Thus, observed discrepancies with predictions found
here may well be due to a such a fundamental shortcoming in the one-zone
models.  In any case, quantitative characterization of the
comparison and contrast between the simplest one-zone models and the
data strongly constrain the formation models.  Here we explore some
possible implications for halo evolution within the one-zone paradigm.

If we take the observed low-metallicity limit of $-4 <$ [Fe/H] at face
value, then this constraint implies that the Simple Model is not a useful
representation of the early Galactic halo evolution.  We have
seen that this low-metallicity threshold is, however, compatible with the
SIMs considered here.  The observed lower limit
effectively sets $\zmin$ for the parent enrichment events, since the
observed metallicity cannot be lower than $\zmin$, excepting
Population~III stars.  We saw above that this constraint on $f(z_0)$
generally yields SIMs for which $nQ \sim 50$ to 100 to fit the
observed halo metallicities, i.e., $n\sim 50 - 100$ generations of
star formation, for contamination filling factor $Q$ of unity.  
Interestingly, this halo $nQ$ is compatible with solar metallicity systems
requiring roughly a Hubble time.  

It seems most plausible that early halo chemical evolution proceeded
inhomogeneously, driven by discrete enrichment events in a warm,
hard-to-cool ISM with little SN-driven turbulence.  This is especially
suggested by the SIM evolutionary sequence seen in the kinematic
subsets shown in Figure~\ref{haloevol}.  However, we have
seen that, above the low-metallicity threshold, the Simple Model does
match the observed halo MDF as well as the low-evolution SIMs
(Figure~\ref{fhalomods}).  Thus it may
be that the early evolution took place inhomogeneously, but rapidly
became essentially homogeneous.  As demonstrated in 
\S 4.3, mixing takes place rapidly if it can take place in a fully
developed HIM.  For starburst-like star formation rates, the HIM
dominates the ISM, thus presumably driving instantaneous mixing,
although on the other hand, metals are then likely to be lost through
galactic outflows. 

For halo evolution dominated by inhomogeneous effects, apparently
a large range in $z_0$ is required, roughly
$-4.7 < \log z_0 < -2$.  The lower limit is set by the observed 
threshold, and the upper limit is constrained by the ability to
reproduce the observed high-metallicity tail.  It is unclear whether
this large range in $z_0$ is unrealistic; as shown in \S 4.3, mixing
processes that dilute the metals and reduce $z_0$ are uncertain.  
Nor do we know whether the mixing processes apply to all, or just
to some, enrichment events.  It does not seem obviously implausible
that the parent enrichment events could span almost three orders of
magnitude in $z_0$, although the range does seem large.  To match the
halo data, the only real requirement for the low-evolution
SIM is that $f(z_0)$ span this range and follow a
decreasing power-law that can fit the high-metallicity tail. 

The above interpretations for halo evolution depend on the 
assumption that the observed limit $-4 <$ [Fe/H] represents a
true threshold.  Should this not be the case, then this would either
imply that homogeneous evolution more strongly dominates, or that the
range in parent $z_0$ is even larger than implied here.  Ultimately,
as discussed above, it will probably be necessary to invoke a more
complex evolutionary process than 
described by these simple one-zone models.  In particular, if the halo is
assembled from a multitude of individual, smaller star-formation
units as suggested in the classic Searle \& Zinn (1978) scenario,
then the coherent formation scenario described by
these simple models presumably would be easily disrupted.

It is also essential to bear in mind that the discrepancy in the predicted
fraction $\fiii$ of Population~III stars poses a profound problem
for all these models.  A variety of solutions has been suggested to
explain the absence of observed Population~III stars, for example,
a primordial IMF composed of high-mass stars, pre-enrichment of the
Galactic halo system, residence of Population~III stars at large
radial distances, disguising of Population~III stars by metal accretion or 
surface convection, etc.  The myriad possibilities are addressed
in the excellent conference proceedings on the first stars (Weiss
{\etal}2000).  Many of these suggestions are compatible with the
simple models discussed above, but again, it may be that the
discrepancy in $\fiii$ is symptomatic of a more fundamental 
omission in these simple models of the halo evolution.  

\section{Conclusion}

We have considered simple, one-zone models for homogeneous and inhomogeneous
chemical evolution of the Galactic halo to estimate the expected fraction
$\fiii$ of zero-metallicity, Population~III stars, and to estimate the
expected low-metallicity thresholds above zero metallicity.  Since the most
metal-poor enrichment events depend on metal dispersal in the ISM, we
also investigated the interstellar mixing and dispersal processes.
We then investigated the results in terms of the observed halo MDF,
offering some possible constraints on the halo evolution.  Interestingly,
the observed halo MDF does not itself preclude a monolithic halo
formation model.  Kinematic subsets of stars previously identified as
``high halo'' and ``low halo'' are broadly consistent with an
evolutionary sequence in terms of the Simple Inhomogeneous Model.

Our mass transport analysis finds that diffusion is inefficient for
metal dispersal since, for the cold and warm ISM phases, the relevant
diffusion lengths are 2 -- 4 orders of magnitude smaller than comparable
turbulent mixing length scales.  This result confirms the earlier
approximations by Tenorio-Tagle (1996), although with some quantitative
differences.  We also find that diffusion is ineffective even in the
HIM, for which the relevant diffusion length is still over an order of
magnitude smaller than the HIM turbulent mixing scale.
Turbulence also likely dominates cooling processes for hot gas.
We estimate some rough relations for the dispersal of metals in the
cold, warm, and hot ISM phases.  As with diffusion, mixing takes place
much more rapidly in hotter phases than cooler phases, assuming that
turbulent velocities are linked to the soundspeed.  We estimate
roughly that turbulent mixing dilutes the metallicity of regions
originating as SN-driven superbubbles by up to an order of magnitude
by volume for ordinary, multiphase ISM. 

This result implies that parent enrichment events may have
metallicities $z_0$ that are up to an order of magnitude lower than in
the absence of interstellar mixing.  Compared to the inhomogeneous
halo model shown in Paper~I, this revises the lowest expected values
of $z_0$ to values that are similar to the lowest observed
metallicities, $-4 <$ [Fe/H].  However, assuming a range in $z_0$ of
about two orders of magnitude, with $\barzero = -4.1$, the resulting
Simple Inhomogeneous Model must be in a highly evolved 
state to show the observed halo metallicities.  The
corresponding MDF drops off too steeply to match the data in the
high-metallicity tail.  This model predicts $\fiii\sim 3\times 10^{-2}$.

However, as found in Paper~I, the observations do match a low-evolution
inhomogeneous model MDF, especially if the range in $z_0$ is extended
to cover almost three orders of magnitude.  This range in values may seem
implausibly large, but the resulting low-evolution model does best
match the observed MDF from among the models considered here.  The
required extension in $\zmin$ could be caused by a number of effects, for
example, an overestimated SN yield and dispersal processes.  This
model predicts $\fiii\sim 4\times 10^{-2}$. 

The homogeneous, Simple Model also matches the observed MDF well,
although it naturally predicts no low-metallicity threshold, and
predicts a fraction $5\times 10^{-3}$ of stars to have [Fe/H] $<-4$,
below the observational limit.  If the observed threshold is real, it
requires a different, presumably inhomogeneous, model to describe the
early halo evolution, since large dispersions in metal abundances are
seen at the lowest metallicities.  It may be possible that the
evolution proceeded more homogeneously at later times.  

Our estimates of $\fiii$ with these simple models are all at least 1
-- 3 orders of magnitude higher than the empirical upper limit of
$\fiii < 4\times 
10^{-4}$.  Given that the observations disagree with both limiting
cases, the homogeneous Simple Model and the Simple Inhomogeneous
Models, we therefore demonstrate an unambiguous discrepancy in the
expected number of Population~III stars.  This poses a significant
problem for these models, although they remain compatible with some of
the proposed solutions.  Recall that this analysis examines only the
most fundamental processes, and does not include parameters like
inflow/outflow and time evolution in $Q$, yield, or IMF, any of which
could have important effects.  However, the constraints offered here 
form the foundation for subsequently exploring such effects.

It is interesting that the current observed lower limit of $-4 <$ [Fe/H]
is consistent with the lowest-metallicity estimates crudely expected for the
parent enrichment events.  Since the lowest metallicities depend on
dilution by interstellar dispersal, in principle the metal-poor regime
of the MDF should also provide constraints on the earliest enrichment
processes.  Our interpretation of the models depends on 
whether the observed limit is indeed real.  Twenty years ago we had an
apparent discrepancy between the expected and observed numbers of
stars having [Fe/H] $< -2.5$; these stars have since been found.
Today, the Simple Model predicts a discrepancy of similar magnitude for
[Fe/H] $< -4$, while the inhomogeneous model does predict this
metal-poor threshold.  Will these additional ``missing'' stars also be
found?  It is essential that the low-metallicity MDF and
threshold, if any, be firmly established.

\section*{Acknowledgments}

It is a pleasure to thank Richard Larson, Anne Sansom, Sergey Silich,
Guillermo Tenorio-Tagle, and especially the referee, Tim Beers, for
comments on the manuscript.  I am also pleased to acknowledge Deidre
Hunter, John Laird, Mordecai Mac Low, and Andy McWilliam for useful
discussions.  Many thanks to John Laird for access to the observed halo
data.   Some of this work was carried out while holding an Institute
Fellowship of the Space Telescope Science Institute, and while
enjoying the hospitality of the Australia Telescope National Facility.

\label{lastpage}

\end{document}